# Updating Maxwell
# with Electrons, Charge, and More Realistic Polarization

*Version 7*

Robert S. Eisenberg

Department of Applied Mathematics

Illinois Institute of Technology

Department of Physiology and Biophysics

Rush University Medical Center

Chicago IL

USA

*Bob.Eisenberg@gmail.com*

March 2, 2020

*File name: Updating Maxwell with Electrons and Charge Version 6 Aug 28-1.docx*

# Abstract

Maxwell's equations describe the relation of charge and electric force almost perfectly even though electrons and permanent charge were not in his equations, as he wrote them. For Maxwell, all charge depended on the electric field. Charge was induced and polarization was described by a single dielectric constant.

Electrons, permanent charge, and polarization are important when matter is involved. Polarization of matter cannot be described by a single dielectric constant $\varepsilon_r$ with reasonable realism today when applications involve $10^{-10}$ sec. Only vacuum is well described by a single dielectric constant $\varepsilon_0$.

Here, Maxwell's equations are rewritten to include permanent charge and any type of polarization. Rewriting is in one sense petty, and in another sense profound, in either case presumptuous. Either petty or profound, rewriting confirms the legitimacy of electrodynamics that includes permanent charge and realistic polarization. One cannot be sure that a theory of electrodynamics without electrons or (permanent, field independent) charge (like Maxwell's equations as he wrote them) would be legitimate or not. After all a theory cannot calculate the fields produced by charges (for example electrons) that are not in the theory at all!

After updating,

(1) Maxwell's equations seem universal and exact.
(2) Polarization must be described explicitly to use Maxwell's equations in applications.
(3) Conservation of total current (including $\varepsilon_0 \, \partial \mathbf{E}/\partial t$) becomes exact, independent of matter, allowing precise definition of electromotive force EMF in circuits.
(4) Kirchhoff's current law becomes as exact as Maxwell's equations themselves.
(5) Conservation of total current needs to be satisfied in a wide variety of systems where it has not traditionally received much attention.
(6) Classical chemical kinetics is seen to need revision to conserve current.

**Introduction.** Maxwell's equations describe the relation of charge and electric forces almost perfectly [1]. Yet Maxwell's equations (as written by Maxwell and Maxwellians before 1897 [2-7]) do not include the electron and its charge. For Maxwell, charge always depended on the electric field. Charge was induced. He did not know of charge that was independent of potential or the electric field [2, 3, 7], p. 36 of [1].

Permanent charge—fixed in magnitude, independent of the electric field—was not part of electrodynamics (for Maxwellians, mostly in the UK, many at Trinity Cambridge) until J.J. Thomson discovered the charge on the electron in 1897 [8]. Just four years earlier, Thomson had described electrodynamics without charge [4]. Other physicists had other opinions, including Weber, Helmholtz, Neumann, Clausius, and Kirchhoff [5-7, 9].

The question is why can traditional Maxwell equations without electrons or permanent charges correctly describe an electrical world with electrons, even though that electrical world contains many charges that are permanent and do not depend on the electric field?

The answer is that the Maxwell equations correctly describe the *relation* of field to charge. Where the properties of charge itself are involved, the equations are incomplete and misleading, and in that sense incorrect[1] [10, 11].

The properties and nature of charge are crucial in important applications, for example, those involving ions in ionic solutions. Almost all biology [12-15] and electrochemistry [16-32] occurs in ionic solutions, along with much of biochemistry and classical chemistry. Ionic solutions consist of permanent charges—the ions but also perhaps in the solvent molecule itself—and so Maxwell's classical equations cannot deal with them, since those equations do not include permanent charges. Maxwell's original equations cannot deal with the role of ions in biology or electrochemistry.

---

[1] If the word 'incorrect' seems too harsh, see p. 4 to read similar characterizations by two Nobel Laureates in Physics. How else would one characterize a theory of gravitation that did not include mass? How else would one characterize a theory of gravitation that dealt with the induced motion ('polarization') of the oceans, as revealed by tides [10,11], and did not deal with the mass of the earth itself?

**Maxwell's equations use a severely oversimplified description of polarization.** Maxwell's equations, as originally written, also have difficulty with the polarization (i.e., dielectric) charge in matter that is induced by electric fields. When an electric field is applied, local forces are large enough to distort distributions of charge within atoms and molecules or perturb them in other ways, e.g., by rotating molecules with asymmetrical permanent charge like water, $H_2O = H^{+0.5\delta} O^{-1.0\delta} H^{0.5\delta}$; $\delta \cong 0.4$. Permanent charge moves in an electric field and some of that movement is called polarization. When local electric fields are zero, polarization is zero, in our use of that word.

Maxwell's equations describe induced charge and polarization with a single dielectric constant. They use one real number $\varepsilon_r \geq 1$. Only the polarization and dielectric properties of a vacuum can be realistically described by a dielectric constant that is a single real positive number. Only in a vacuum does $\varepsilon_r = 1$.

Polarization charge in real materials varies dramatically with time. In our electronic devices, time scales range from $10^{-10}$ to $> 10^0$ sec [33-50]. Polarization in ionic solutions varies by a factor of ~40 in the time range linking atomic motions to macroscopic movements $10^{-15}$ to $> 10^0$ sec, estimated by the effective dielectric coefficient measured in innumerable experiments. This time scale is used in hundreds of simulations of molecular dynamics of proteins every day. Linkage of atomic motion and macroscopic function is a central issue in molecular biology and electrochemistry because atomic structure controls macroscopic function in life and technology. References [17, 22, 23, 27, 51-60] provide documentation of the properties of the effective dielectric constant.

Polarization and other charge movements are also produced by fields not included in the Maxwell equations at all, and thus not described by them, at all. Many forces beyond the electrical move charge in ionic solutions and these are involved in a wide variety of applications. Convection (as in a garden hose filled with saltwater) and diffusion (as in nearly every biological or electrochemical system) are examples. Biology, chemistry, engineering, and physics all have electric current driven by forces not contemplated by or included in Maxwell's original equations.

Dealing with multiple types of forces and fields is tricky, even when only linear dielectrics are involved: the Abraham-Minkowski paradox continues [61-65] and may not have been resolved to the satisfaction of everyone [66, 67] despite the attempts at simplification by one of the most brilliant and successful innovators in the history of physics [62].

When multiple types of forces and field are involved, charge is transported on particles that move. Then, all the flows of fluid mechanics [68-70] are involved and must be dealt with by theory, individually and in their interactions with each other and the electric field.

Charged particles also can change shape as they move, thereby changing spatial distributions of permanent and induced charge. They dissipate energy because charged structures contain atoms that collide (and produce friction) as they move. They form elastic structures that store energy in the electric field as they move.

The theory of complex fluids [71-81] is then needed and useful. Conservation of mass must be combined with electrodynamics in these cases. All of electrodynamics is not always involved. Conservation of total current is enough in the special one dimensional systems that are the electronic circuits of our digital technology, see eq. (28); Maxwell's equations and its **B** field, see eq. (26) & (27) are needed in general systems.

The theory of complex fluids can be combined with the Maxwell equations in an energetic variational approach [72, 74, 76, 82-85] that has dealt successfully with electro-magneto-hydrodynamics flows [52], liquid crystals, polymeric fluids [86, 87], colloids and suspensions [79, 88] and electrorheological fluids [52, 89]. Variational methods describe solid balls in liquids; deformable electrolyte droplets that fission and fuse [79, 90]; and suspensions of ellipsoids, including the interfacial properties of these complex mixtures, such as surface tension and the Marangoni effects of 'oil on water' and 'tears of wine' [79, 88, 91, 92].

These situations, where Maxwell's equations are incomplete, and in that sense incorrect, are not exotic. Even flowing seawater cannot be described by Maxwell's equations, as he knew them. Fields like diffusion and convection move and induce charge that cannot be described by Maxwell's equations, let alone with polarization described by a single dielectric coefficient $\varepsilon_r \geq 1$. A search for the literature of Poisson-Nernst-Planck equations (PNP [93-100] or drift diffusion [39, 43, 51, 101-106] are other names) will show that the issues and applications of coupled diffusion and electrodynamics cover a wide span of science, technology, and life, from cement to biological calcium channels, reported in hundreds of papers.

Astrophysics [107-112] adds heat flows and temperature fields to that list. Nuclear physics, from power plants to thermonuclear weapons, involve still other fields of importance to our lives, judging by the funds spent on them. Applications as diverse as desalination, detoxification, super capacitors, and even cement technology [113], require the treatment of multiple fields. Important applications, particularly optics [114-117], involve induced charge that has nonlinear dependence on electric fields not comfortably described by a single dielectric constant.

None of this is meant to take away from the immense contributions of Maxwell [118, 119]. Indeed, I know of no other scientific contribution which has (quantitatively) stood the test of time—154 years of criticism—as well, and as accurately, as the Maxwell equations, suggesting that Maxwell was a very special scientist indeed.

Maxwell's equations can be misleading, however, as special as they are, when applied to matter, because matter includes permanent charge and types of induced polarization not known to Maxwell and his contemporaries. As brilliant scientists as they were, the Maxwellians could not know what they did not know, like the rest of us. It is wonderful that their reach exceeded their grasp, but their reach did not touch [2-4, 6], let alone grasp the electron and its permanent charge, or the movements of charge driven by non-electric fields, constrained by conservation of mass or the diverse movements of charge that can create polarization.

**Historical note.** Others (including the two Nobel Laureates in Physics I quote in the next two paragraphs) have views like those just presented. But stating these views seems radical to many, and generates skepticism, even disbelief, in my experience. It seems clear that our enormous respect for the contributions of Maxwell can make the realities of polarization and charge hard to discuss.

Feynman's text says (on p. 10-7 of [120]) "One more point should be emphasized. An equation like $\mathbf{D} = \varepsilon_r \varepsilon_0 \mathbf{E}$ [our eq. (2)] is an attempt to describe a property of matter. But matter is extremely complicated, and [our] eq. (2) is in fact not correct." Feynman describes how the equation is not correct in some detail in four sentences that list many of the points made in pages 5-7 of this paper. Feynman then says "It [our eq. (2)] cannot be a deep and fundamental equation."

Purcell and Morin Ch. 10 of [121] states "the introduction of **D** is an artifice that is not, on the whole, very helpful. We have mentioned **D** because it is hallowed by tradition, beginning with Maxwell, and the student is sure to encounter it in other books, many of which treat it with more respect than it deserves." Purcell and Morin [121] also states[2] on p. 506, quite remarkably, in my view, "the distinction between bound and free charge is ambiguous." and then on p. 507 "This example teaches us that in the real atomic world the distinction between bound charge and free charge is more or less arbitrary, and so, therefore, is the concept of polarization density **P**."

The significance of the quoted statements is hard to overstate. If eq. (2) is in fact 'not correct'; if the concept of polarization is 'more or less arbitrary'; and the distinction between bound and free charge is 'ambiguous', then the formulation of the Maxwell equations in [120, 121] and many other textbooks is ambiguous and arbitrary. I agree with the opinion of ref [65] that " …. the conventional theory of electrodynamics inside matter needs to be redesigned".

Feynman's redesign [120] is "… to exhibit in every case all the charges, whatever their origin, [so] the equations are always correct.". The implication is that nothing much can be done unless the properties of all the charges, whatever their origin, are known in detail. That was my view from 1963 when I first read Feynman's words, thanks to my Harvard roommate and friend physicist Peter Koehler. I suspect that view is shared widely by workers on electrodynamics. Hence, my surprise when the updating of Maxwell's equations showed that Feynman's view was incorrect. In fact, a very important corollary of Maxwell's equations can be derived without any mention of matter whatsoever. The conservation of current eq. (25) is true **without any discussion of charges.**

This paper is meant to show another way to do the redesign and create the fundamental equations Feynman was discussing. Our redesign is nearly trivial, but in one respect (in my immodest opinion) the redesign is profound. Our redesign derives a crucial equation (25) that is valid even when 'matter is very complicated' (p. 10-7 of [120]). The extended discussion of these issues, here and elsewhere [60, 122-125], was motivated by my surprise that Feynman was incorrect. I had the feeling that my surprise would be shared by many others, albeit after suitable skepticism.

---

[2] I paraphrase for clarity.

To summarize, ***Maxwell's equations require rewriting***

(a) to include permanent charge

(b) to accommodate the actual properties of polarization and induced charge

(c) to include flows driven by other forces.

Other rewrites may be necessary involving issues I do not know or understand [65, 126, 127]. This paper is surely incomplete.

Rewriting might be thought to be a petty, nearly trivial academic exercise, more tedious than profound, surely presumptuous. Those thoughts may be why Maxwell's equations have not been rewritten (as proposed here) in standard textbooks by authors who certainly knew of the complexity of permanent charge long before I did.

Rewriting the Maxwell equations turns out, however, to be more than an academic exercise.

(1) Rewriting shows where Maxwell's original restricted treatment of charge limits understanding and leads to misunderstandings.

(2) Rewriting makes it easier to extend electrodynamics to forms of charge and charge movement driven by other forces and fields, like convection, diffusion, heat flow and so on.

(3) Rewriting shows that it is no longer necessary to separate the treatment of macroscopic and microscopic (atomic) scale electrodynamics[3], as is done in many textbooks. Because all types of charge, flux of charge, and current are treated explicitly in the rewriting, one treatment of Maxwell's equations is enough.

---

[3] What are called microscopic equations in many textbooks actually refer to an atomic scale, far below the resolution of light microscopes.

(4) Rewriting focuses attention on the universal legitimacy of Maxwell's equations. The utility, and thus legitimacy of the Maxwell equations becomes clear if all types of charge and flow are treated explicitly without imbedded or implicit approximations to any properties of matter.

In fact, **Maxwell's equations as originally written are not legitimate**, for present day applications, because they leave out the properties of electrons and ions, both enormously important sources of the electric field.

Even the standard textbook treatments of Maxwell's equations have limited legitimacy because they do not apply to the common situation of present-day applications where polarization cannot be described by a single dielectric constant.

The standard textbook treatments of Maxwell's equations have limited legitimacy because they do not describe situations where charge moves and electric fields are created by other fields and forces not mentioned in Maxwell's equations at all, such as diffusion and convection. Diffusion and convection nearly always occur in biological and electrochemical applications. Those are two of the most important applications of electrodynamics.

Scientists concerned with atomic scale phenomena—chemists, biochemists, and biologists—have questioned how useful equations of electrodynamics can be if they leave out all forces and fields other than electromagnetism and if they depend on the drastically inappropriate approximation of a single dielectric constant. The answer to the question "How useful are the equations of electrodynamics?" comes from rewriting Maxwell's equations, in my opinion.

Rewriting confirms the legitimacy of Maxwell's equations and the need to use them on atomic as well as macroscopic scales.

(5) Rewriting exposes a misunderstanding of the nature of charge in ionic (and protein) solutions [128-135]. Protein solutions are particularly important because the blood, plasma, and insides of biological cells are intracellular protein solutions. Proteins function in ionic solutions. Proteins are enormously important in biology because they are "life's robots" [136] that perform most of life's functions as they form many of its structures.

Classical chemistry describes substances by potential surfaces, and the rate constants they support, but our rewritten electrodynamics (Appendix of [137]) shows that almost anything dissolved in a solution should be described as a surface of permanent charge (as a first approximation, neglecting polarization at the surface), not a surface of potential [128, 129]. Polarization can produce a range of devices [138-140] but our focus of attention is on the first order effects of permanent charge on proteins, analogous to doping in semiconductor devices [38, 39, 43, 104, 105].

(6) Rewriting allows deeper understanding of Faraday's 'electromotive force' EMF [7] that moves charge and creates current in electric circuits. Specifically, Maxwell's equations for electric circuits can now be solved for $\mathbf{E}(x, y, z|t)$ in eq. (29) giving a modern definition of an idea—EMF of circuits—that previously might have seemed mysterious or vague to some of us.

$\mathbf{E}(x, y, z|t)$ provides the pondermotive force−the EMF of circuits−that

(6a) moves charges with mass $\mathbf{Q}_{...}(x, y, z|t; \mathrm{E})$.
(6b) creates the material currents $\mathbf{J}_{...}(x, y, z|t; \mathrm{E})$.
(6c) helps satisfy conservation of total current $\mathbf{J}_{total}$ in eq. (25).

(7) Most importantly, rewriting shows that conservation of total current is an exact and universal law, independent of the properties of matter altogether, if current is defined, as in eq. (24), to include the polarization of space (i.e., of the vacuum, defined in eq. (4)) that is the ethereal current that flows in a vacuum. It is interesting that Maxwell defined total current as we do, according to his followers at Trinity Cambridge, Jeans and Whittaker: ref. [141], Ch. 17, p.511; [7], p. 280, respectively.

(7a) Conservation of current is usually derived using formulations of Maxwell's equations that include a single dielectric constant. That approximation is so unrealistic [59] that the utility of Maxwell's equations as applied to matter comes into question.

**Conservation of current then seems illegitimate, not universal, because its derivation uses an illegitimate approximation.**

Legitimacy is restored when conservation of total current is derived without reference to matter, entirely independent of dielectric properties. Conservation of current is then a universal law (eq. (25)), valid inside atoms [60], wherever the Bohm version of quantum mechanics can be applied [142-146]. Conservation of total current is seen to be as exact and universal as the Maxwell equations themselves.

(7b) With the definition of total current, Kirchhoff's current law for electrical circuits (i.e., branched one dimensional systems) becomes exact [125, 147] and we can understand how Kirchhoff's law can serve as the main theoretical tool used by the engineers who design our digital technology. That technology operates at times shorter than $10^{-9}$ sec [35, 46, 47, 148-151]. Remember that light travels approximately one foot in one nanosecond, so the validity of Kirchhoff's law—like the performance of electronic circuits at these speeds—is a welcome surprise. (Textbook treatments of Kirchhoff's law deal with the steady state. Derivations treat Kirchhoff's law as approximate, limited to relatively long times [152-155]. Current at the $10^{-9}$ sec time scale is certainly not steady state or on a long-time scale. Current on that time scale is not just the movement of charges with mass, as an experimental fact described in detail in [124] and other references cited there.)

(7c) With this definition of total current, the equations of chemical kinetics do not satisfy conservation of current, until modified. A series of chemical reactions described by the law of mass action do not all have equal currents as required by the conservation law. Standard chemical kinetic models need to be modified so a series of them conserve current. Perhaps a network that satisfies conservation of current needs to be solved along with the classical network that satisfies conservation of mass.

**Motivation.** These issues (1)–(7) are discussed at such length in this paper and its predecessors [60, 122-125] because in my opinion they are pivotal, of great general importance to science and technology. The flow of current in chemical reactions like the electron chains of oxidative phosphorylation and photosynthesis [156] require an updating of Maxwell's equations. The theory of chemical reactions of charged reactants needs to be revised to conserve current, in my opinion.

These issues underly the design process of the circuits of our digital technology and have thus catalyzed its success, in my opinion. These issues are the reason that our electronic circuits have increased in capability by a factor of nearly $10^9$ in our lifetimes. The unprecedented success of that electronic technology has transformed the nature of human life by allowing the interactions of village life to extend to the family of man with its nearly $10^{10}$ members.

My fantasy is that exploiting the universal nature of conservation of current could also transform the technology of ions in biological and electrochemical systems. Or, more modestly, one might say that our understanding and development of electrochemical systems cannot fully develop until polarization is correctly described and electrodynamics is written in modern form.

**Other redesigns**. The redesign presented here is minimal. It was chosen to make the smallest changes possible to deal with permanent charge, driven by nonelectrical forces, and polarization of any type. It is designed to look like the standard treatment of textbooks. The redesign is limited because it uses the vector calculus. It avoids issues of special and general relativity important for some applications, that involve generalizations of vector calculus, beyond my scope. Our redesign is surely incomplete.

Issues of special and general relativity are dealt with in ref [65] using exterior differential forms. Despite its treatment of relativity, ref [65]—and other modern references that I know of [126, 127]—do not deal explicitly with the issues enumerated on p. 8-11.

One must realize that special relativity is not general enough even in ordinary systems: special relativity does not deal with rotational motion and the accelerations involved. Rotations certainly occur in the range of applications of interest here. Gravitation seems to be involved in only the most classical Newtonian sense in our systems so perhaps a mild generalization of special relativity, dealing with rotational motion but not gravitation, would be enough for our purposes, provided such a treatment is possible and known.

A useful redesign of electrodynamics must deal with the complexities of polarization, permanent charges, and charge movement, in my opinion. Otherwise, electrodynamics is not general enough to deal well with ionic solutions, and thus with much of chemistry, and most of electrochemistry, biochemistry, and biology, no matter how well the redesign deals with relativity, special and general.

## **Theory and Derivation**

We begin the derivation section with the modern textbook presentation of Maxwell's first equation, showing how it can accommodate what we now know of matter and charge, that was not known when Maxwell wrote these equations. We choose this approach so this paper is accessible to scientists who have learned electrodynamics from textbooks of the last hundred years or so, following [157, 158].

We start with

### **Maxwell's First Law**

$$\mathbf{div}\, \mathbf{D}(x,y,z|t) = \boldsymbol{\rho}_f(x,y,z|t) \qquad (1)$$

The divergence operator is described usefully in [159-161]. Eq. (1) is a classical form of Maxwell's electrostatics equation, Maxwell's first law, where $\boldsymbol{\rho}_f(x,y,z|t)$ is free charge, identified in eq. (12) & (13) as $\boldsymbol{\rho}_f(x,y,z|t) = \boldsymbol{Q}_{perm} + \boldsymbol{Q}_{other}$.

The dielectric properties of matter are combined with the electric field $\mathbf{E}(x,y,z|t)$ in the auxiliary displacement field $\mathbf{D}(x,y,z|t)$ using the relative (dimensionless) dielectric constant which here is a single real positive number $\varepsilon_r \geq 1$. We follow the language and approach of a textbook [1], Ch. 3 (p. 167, eq. 6.36) for the convenience of the reader, using $\varepsilon_0$ for the electrical constant, the permittivity of the vacuum.

$$\mathbf{D}(x,y,z|t) = \varepsilon_r\, \varepsilon_0 \mathbf{E}(x,y,z|t) \tag{2}$$

Another classical form of Maxwell's first law, assuming no spatial or other dependence for the dielectric constant $\varepsilon_r$ is

### Maxwell's First Law

$$\varepsilon_r \varepsilon_0 \mathbf{div}\, \mathbf{E}(x,y,z|t) = \boldsymbol{\rho}_f(x,y,z|t) \tag{3}$$

The dielectric constant $\varepsilon_r$ is a single real number in this statement.

It is necessary to state the obvious, to avoid confusion: Maxwell's first law is useful when the properties of $\boldsymbol{\rho}_f$ can be specified independent of the law itself. In applications, Maxwell's equations are coupled to other equations that describe $\boldsymbol{\rho}_f$ or to tables of data that specify $\boldsymbol{\rho}_f$ from experiments. Only currents determined by eq. (25) can be specified without such knowledge.

It is necessary also to reiterate that $\varepsilon_r$ is a single, real positive constant in Maxwell's equations as he wrote them and as they have been stated in many textbooks since then, following [141, 157, 158]. If one wishes to generalize $\varepsilon_r$ so that it more realistically describes the properties of matter, one must actually change the differential equation (3) and the set of Maxwell's equations as a whole. If, to cite a common (but not universal) example, $\varepsilon_r$ is to be generalized to a time dependent function (because polarization current in this case is a time dependent solution of a linear, often constant coefficient, differential equation that depends only on the local electric field), the mathematical structure of Maxwell's equations changes.

Solving the equations with a constant $\varepsilon_r$ and then letting $\varepsilon_r$ become a function of time creates a mathematical chimera that is not correct. The chimera is not a solution of the equations. Even if one confines oneself to sinusoidal systems (as in classical impedance or dielectric spectroscopy [54, 55, 57, 162]), one should explicitly introduce the sinusoids into the equations and not just assume that the simplified treatment of sinusoids in elementary circuit theory [46, 48, 163-165] is correct: it is not at all clear that Maxwell's equations—combined with other field equations (like Navier Stokes [76, 84, 85, 97, 166-179] or PNP = drift diffusion [39, 43, 51, 93-106]); combined with constitutive equations; and boundary conditions—always have steady state solutions in the sinusoidal case. They certainly do not always have solutions that are linear functions of just the electric field [114-117].

# Polarization

Polarization describes two quite different kinds of physics, as the word is commonly used.

## Vacuum displacement or vacuum polarization

$$\text{Polarization of space is the } \textbf{displacement current} = \varepsilon_0 \frac{\partial \mathbf{E}}{\partial t} \quad (4)$$

Polarization is a universal property of space, whether in a vacuum or filled with matter, that allows light to propagate in a vacuum even though the vacuum contains no charge with mass. The polarization of space $\varepsilon_0\ \partial\mathbf{E}/\partial t$ is hard to understand in the mechanical systems used by Maxwell early in his career, and so might be called an ethereal current $= \varepsilon_0\ \partial\mathbf{E}/\partial t$.

**The ethereal current** $\varepsilon_0\ \partial\mathbf{E}/\partial t$ arises from the invariance of charge with velocity, even velocities approaching the speed of light. Unlike mass, length (or other dimensions), and time, charge does not change as velocities approach the speed of light. The displacement current term (and its presence in Ampere's law, in Maxwell's version, eq. (14)) are consequences of these facts. The relation of Maxwell's equations and special relativity are beyond my scope and thus the scope of this paper. The subject is covered in detail in many texts, e.g., [48, 65, 126, 127, 180-183]. In fact, special relativity is not general enough because it is confined to inertial systems. It does not deal with (non-inertial) rotating systems and those cannot be ignored in practical applications, including trying to understand what 'spin' of an electron might mean in an electrodynamic, nonquantum manner of thought.

Interesting questions about spin and the Pauli exclusion principle arise even on the macroscopic scale: How does the trajectory of a pair of electrons differ from that of negatively charged magnets (i.e., two separate magnetic point dipoles each with one negative charge that might be called pseudo-electrons)? Do two macroscopic charged magnets pair up with special energy when placed in a radial electrostatic field generated by a single positive charge, analogous to a proton? Do two charged dipole magnets pair up in a special way to the exclusion of other pseudo-electrons, reminiscent of Pauli exclusion on the atomic scale? No matter what energy and special properties the pair of pseudo-electrons might have on the macroscopic scale, quantum mechanics is needed, of course, on the atomic scale.

**Polarization of Matter**. The classical auxiliary polarization field $\mathbf{P}(x, y, z|t)$ is used here, following so many others, to define the part of electric charge that depends linearly on local electric fields. We follow textbooks, e.g., eq. 6.25 and 6.36 of [1] and define polarization $\mathbf{P}(x, y, z|t; \mathbf{E})$ to be a property of matter, but we exclude polarization of the vacuum, because it has such a different origin. Polarization $\varepsilon_0 \, \partial \mathbf{E}/\partial t$ of the vacuum is a property of space. Polarization of eq. (5) $\mathbf{P}(x, y, z|t; \mathbf{E})$ is a property of matter.

## **Polarization of Matter**

$$\mathbf{P}(x, y, z|t; \mathbf{E}) = \mathbf{D}(x, y, z|t) - \varepsilon_0 \mathbf{E}(x, y, z|t) = (\varepsilon_r - 1)\varepsilon_0 \mathbf{E}(x, y, z|t) \tag{5}$$

We define $\mathbf{P}(x, y, z|t; \mathbf{E})$ to be zero when the local electric field is zero. Thus, the following types of charge are not included in the $\mathbf{P}(x, y, z|t; \mathbf{E})$ defined here. They are included in another way, in the permanent charge $\mathbf{Q}_{perm}$ or $\mathbf{Q}_{other}$ of eq. (12).

(1) electrets [184, 185];
(2) macro dipoles of molecules and chemical bonds, e.g., of carbonyl bonds or (perhaps delocalized) carbon oxygen bonds of carboxylic acids [18, 27, 186-188].
(3) the point dipoles of atoms that are present when the electric field is zero are included in another way, in the permanent charge $\mathbf{Q}_{perm}$ of eq. (12).
(4) charge that depends in a nonlinear way on the local electric field is included in the permanent charge $\mathbf{Q}_{other}$ of eq. (12).

We reconcile this traditional usage with the properties of matter and charge as known today by isolating the polarization of just **<u>ideal dielectrics</u>** as $\mathbf{P}_{dielectric}(x,y,z|t;\mathbf{E})$

$$\mathbf{P}_{dielectric}(x,y,z|t;\mathbf{E}) = (\varepsilon_r - 1)\varepsilon_0 \mathbf{E}(x,y,z|t) \tag{6}$$

The ideal dielectric polarization $\mathbf{P}_{dielectric}(\cdots|t;\mathbf{E})$ varies with time only because the electric field $\mathbf{E}(\cdots|t)$ varies with time.

$\mathbf{P}_{dielectric}(x,y,z|t;\mathbf{E})$ is not equal to the polarization $\mathbf{P}(x,y,z|t;\mathbf{E})$ because of the non-ideal properties of matter. Our treatment departs from standard textbooks because we deal explicitly with the difference $\mathbf{P} - \mathbf{P}_{dielectric}$. The difference $\mathbf{P} - \mathbf{P}_{dielectric}$ can have any properties whatsoever. For example, it may depend on fields and forces not written in Maxwell's equations at all, fields like convection or diffusion.

Obviously, the nonideal properties $\mathbf{P} - \mathbf{P}_{dielectric}$ must be known from theory or experiments before Maxwell's equations can be applied to compute forces, flows and energy in practical problems. Only eq. (25) is useful without such knowledge.

$\mathbf{P}(x,y,z|t;\mathbf{E})$ in general, involves the movement of charges with mass. As these move, they are damped by collisions with other particles and suffer dissipation through friction. Friction introduces time dependence beyond that of the electric field itself, and so when friction is significant, variables that do not include friction are obviously inadequate. Then, the idealized $\mathbf{P}_{dielectric}(x,y,z|t;\mathbf{E})$ and eq. (6) are a poor, often strikingly poor approximation to the real properties of matter, including its polarization. It should be emphasized that atomic charges in liquids are masses in a condensed phase [18] which can only move by interacting ('colliding') with other particles and experiencing friction that dissipates their kinetic energy. Thus, the movement of atoms and molecules in condensed phases involves time dependence that cannot be described by the idealized $\mathbf{P}_{dielectric}(x,y,z|t;\mathbf{E})$ and eq. (6). There is little charge movement in biology and chemistry (in liquids or condensed phases [17, 22, 23, 27, 51-58]) that can be well described by the idealized $\mathbf{P}_{dielectric}(x,y,z|t;\mathbf{E})$ and eq. (6).

Permanent dipoles, and other spatial distributions of charge that are independent of the electric field, are not part of the idealized $\mathbf{P}_{dielectric}(x,y,z|t;\mathbf{E})$ as just discussed. Rather, they are included in the permanent charge $\mathbf{Q}_{perm}(x,y,z|t)$ which is defined here to be the

charge that is entirely independent of the electric field. Properties of electrets or chemical bonds that vary with the electric field in the simple way described by eq. (5) are treated here as part of the idealized $\mathbf{P}_{dielectric}(x,y,z|t;\mathbf{E})$. Properties of electrets or chemical bonds that vary in a more complex way with electric field are described by charges $\mathbf{Q}_{other}(x,y,z|t;\mathbf{E})$ and the flux of these charges by $\mathbf{J}_{other}(x,y,z|t;\mathbf{E})$. Those properties include charges and fluxes that depend nonlinearly on the local electric field; charges and fluxes that depend on global properties of the fields—not just the local electric field—and charges and fluxes with complicated time dependence, dependent, for example, on other variables that are themselves specified by their own partial differential equations

This separation may seem unwise because it (unfortunately) splits properties of one physical system into components, the way $kx^2$ might be split from $kx$ in a Taylor expansion of $ke^x$, where $k$ is a constant.[4] It is also unwise to abandon the language and approach of electrodynamics which scientists have learned for some one hundred fifty years. The separation is made to retain classical notation.

**Polarization of real matter** is as complex as the dynamics of matter itself. Detailed properties of polarization are important in chemistry, technology, and biology over an enormous range of time and length scales (say $10^{-20}$ sec to $10^2$ sec and $10^{-11}$ meter to 1 meter and over a much larger range in high energy physics and astrophysics. See [17, 22, 23, 27, 51-58, 120] and documentation cited in [59, 60].

The diverse spectra of atoms and molecules come from the bewilderingly complex properties of polarization [55, 56, 58, 189-193]. Polarization determines the interaction of light (i.e., electromagnetic radiation) and molecules and atoms. Parsegian [194] describes in detail the connection between polarization and spectra. The spectra of molecules are remarkably diverse, almost as diverse as the molecules themselves [190, 191, 193-196]. Spectra are used to identify molecules—the way fingerprints identify people—because they are so sensitive to chemical structure.

---

[4] Scientists trying to estimate the coefficients of $\hat{k}x^2$ and of $\check{k}x$ in a two term approximation to a process $ke^x$ may not realize the two coefficients $\hat{k}$ and $\check{k}$ are equal once the approximation is split into disjoint terms. Scientist need extra information before the coefficients $\hat{k}$ and $\check{k}$ can be set equal.

The spectra—and thus the polarization—of real materials obviously cannot be described by a single dielectric constant.

Polarization is a nearly universal property of atoms and of matter, but the electron itself does not polarize in the ordinary sense of that word. A crucial property of the electron is that its charge is entirely independent of the electric field. Polarization of the electron [197, 198] involves its magnetic properties, called 'spin' and does not involve the size of the charge.

Indeed, the charge on the electron is altogether constant. It is even independent of velocities close to the speed of light, unlike length, mass, and (amazingly enough to me) time itself. These issues are discussed at length in ref [48, 60, 65, 123-127, 180-183].

**Permanent Charge**. Permanent charge was first recognized in electrons in a vacuum. Today we know that charges on many ions (like sodium $Na^+$ and potassium $K^+$) in the electrolyte solutions that are 'the liquid of life' are also constants, independent of the electric field around them. Electrons, sodium, and potassium ions are permanent charges. The charges of the acid and base side chains of proteins have many of these properties as well. These charges are permanent, with a fixed value of charge, although of course they move and so are not fixed in time or space. That is why I prefer the name 'permanent' charge to the more traditional 'fixed' charge.

All of biology [12-15, 199-205], much of chemistry, and a wide range of technological applications involve charged materials that move and diffuse [17, 18, 22, 24, 26, 27, 30, 31, 206-208]. The equations of electrodynamics must be extended to deal with those movements. Electrodynamics must be combined with fluid mechanics (say, the Navier Stokes equations [76, 84, 85, 97, 166-179]) to describe the convection of ionic solutions [76, 84, 85, 97, 166-179]. A suitable description of diffusion must be included in nearly every biological and chemical application (say the Poisson-Nernst-Planck, PNP [93-100] or drift diffusion equations [39, 43, 51, 101-106]). A description of heat flow is needed in some applications as well. These other types of flow satisfy conservation of matter, along with constitutive equations too diverse to summarize in a few equations, let alone words.

The charge involved in these interactions is described by the charge density $\mathbf{Q}_{other}(x, y, z|t; \mathbf{E})$ and its flux by $\mathbf{J}_{other}(x, y, z|t; \mathbf{E})$, used later in this paper.

We now rewrite Maxwell's first equation to separate different kinds of charge depending on how they vary with the electric field.

## **Rewriting Maxwell's Equations**

We start with Maxwell's first equation relating charge and the electric field by writing out all components of charge, permanent charge $\mathbf{Q}_{perm}$; polarization of space represented by $\varepsilon_0$; polarization of matter represented by $(\varepsilon_r - 1)\varepsilon_0$; and every other charge density by $\mathbf{Q}_{other}$.

### **Maxwell's First Equation**

$$\big((\varepsilon_0(\varepsilon_r - 1) + \varepsilon_0)\,\mathbf{div}\,\mathbf{E}(x,y,z|t)\big) = \mathbf{Q}_{perm}(x,y,z|t) + \mathbf{Q}_{other}(x,y,z|t;\mathbf{E}) \tag{7}$$

$$\mathbf{div}\,\varepsilon_0 \mathbf{E}(x,y,z|t) = \mathbf{Q}_{perm}(x,y,z|t) + \mathbf{Q}_{other}(x,y,z|t;\mathbf{E}) - (\varepsilon_r - 1)\varepsilon_0\,\mathbf{div}\,\mathbf{E}(x,y,z|t) \tag{8}$$

From eq. (6)

$$\mathbf{div}\,\varepsilon_0 \mathbf{E}(x,y,z|t) = \mathbf{Q}_{perm}(x,y,z|t) + \mathbf{Q}_{other}(x,y,z|t;\mathbf{E}) - \big(\mathbf{div}\,\mathbf{P}_{dielectric}(x,y,z|t;\mathbf{E})\big) \tag{9}$$

and this could be further rewritten defining the charge of an *idealized* dielectric $\mathbf{Q}_{dielectric}$ created by *idealized* polarization $\mathbf{P}_{dielectric}$, if $\mathbf{Q}_{dielectric}$ is significant.

$$\mathbf{Q}_{dielectric}(x,y,z|t;\mathbf{E}) = -\mathbf{div}\,\mathbf{P}_{dielectric}(x,y,z|t;\mathbf{E}) \tag{10}$$

$$\mathbf{Q}_{dielectric}(x,y,z|t;\mathbf{E}) = (\varepsilon_r - 1)\varepsilon_0\,\mathbf{div}\,\mathbf{E}(x,y,z|t) \tag{11}$$

We reiterate (to avoid possible confusion) that $\mathbf{Q}_{dielectric}$ and $\mathbf{P}_{dielectric}$ describe the properties of ideal dielectrics, not the dielectric properties of real materials.

This gives a particularly useful form of **<u>Maxwell's First Equation</u>**

$$\mathbf{div}\,\varepsilon_0 \mathbf{E}(x,y,z|t) = \mathbf{Q}_{perm}(x,y,z|t) + \mathbf{Q}_{dielectric}(x,y,z|t;\mathbf{E}) + \mathbf{Q}_{other}(x,y,z|t;\mathbf{E}) \quad (12)$$

Eq. (12) seems particularly useful because it separates charge according to its physical properties. $\mathbf{Q}_{perm}(x,y,z|t)$ is independent of the electric field. The permanent charge of the idealized dielectric $\mathbf{Q}_{dielectric}(x,y,z|t;\mathbf{E})$ depends linearly on the magnitude of the electric field as shown in eq. (5) & (6). This term characterizes the polarization of ideal dielectrics by a single real number, a dielectric constant $\varepsilon_r$. $\mathbf{Q}_{other}(x,y,z|t;\mathbf{E})$ describes all other charges. They all have mass and can be moved by forces not ordinarily included in equations of electrodynamics. Examples are convection, diffusion, and heat.

We recognize the classical **<u>free charge</u>** on the right hand side of eq.(12)

$$\boldsymbol{\rho}_f(x,y,z|t;\mathbf{E}) = \mathbf{Q}_{perm}(x,y,z|t) + \mathbf{Q}_{other}(x,y,z|t;\mathbf{E}) \quad (13)$$

found in the classical formulation of eq. (1). The ideal dielectric term $\mathbf{Q}_{dielectric}$ is not seen in eq. (13) because it is described by the dielectric constant $\varepsilon_r$ in eq. (11). Permanent charge arising from nonideal properties of polarization is described by $\mathbf{Q}_{other}$, if it exists.

We again point out that Maxwell's equations become a useful tool to compute forces, flows, and energy in practical problems only if the properties of $\boldsymbol{\rho}_f$, $\mathbf{Q}_{perm}$ and $\mathbf{Q}_{other}$ are specified. The specification might be experimental data itself, in a massive look up table. More commonly, the specification is another set of differential-integro field equations, called constitutive relations, that need to be solved together with Maxwell's equations. Only currents specified by eq. (25) can be determined without such knowledge.

**<u>Complex fluids.</u>** $\boldsymbol{\rho}_f$, $\mathbf{Q}_{perm}$ and $\mathbf{Q}_{other}$ can involve all the properties of matter and its movement, so electrodynamics does not provide a complete description. Other fields are involved, like convection and diffusion.

The techniques of the theory of complex fluids [71-81], and its energetic variational calculus [72, 74, 76, 79, 82-85, 90, 170, 209, 210], help ensure that sets of field equations are satisfied consistently, with all variables satisfying all equations and boundary conditions, with one set of parameters.

**Non-transferable theories**, like much of chemical kinetics, often use parameters that vary with experimental conditions in ways not predicted or understood by theory.

The importance of using one set of parameters may seem too obvious to mention. However, some fields of science use 'nontransferable parameters' to ensure that a particularly favored equation fits data as the equation is transferred from one set of experimental conditions to another [134, 135]. Chemical kinetics uses the law of mass action this way. As admirable as it is to be sure equations always fit data, it is difficult to design a stable, robust transferrable technology if it is built on a shifting, nontransferable foundation.

It might seem a fool's errand, treading on an angel's insights (Feynman [120]) to try to say anything general about properties of electric fields in eq.(12) " … without exhibit[ing] in every case all the charges, whatever their origin, …" (p. 10-7 of [120]). But a fool can rush in usefully without knowing the origin and properties of charges, saying something general and exact about current flow, as we shall see in eq.(25).

Something general can be said—without specifying constitutive field equations—because of the properties of the fluxes $\mathbf{J}_{...}$ included in what we might call Maxwell's second equation, his version of Ampere's law, that describes the magnetic field $\mathbf{B}(x, y, z|t; \mathbf{E})$.

**Maxwell's Second Equation, his Ampere's Law**

$$\boxed{\frac{1}{\mu_0} \text{curl } \mathbf{B} = \mathbf{J}_{dielectric} + \mathbf{J}_{permanent} + \mathbf{J}_{other} + \varepsilon_0 \frac{\partial \mathbf{E}}{\partial t}} \qquad (14)$$

Each of the fluxes $\mathbf{J}_{...}$ can depend on location and time and electric field, as functions of $(x, y, z|t; \mathbf{E})$ but the notation is condensed for clarity. $\varepsilon_0 \, \partial \mathbf{E}/\partial t$ is the displacement current present everywhere, that describes the polarization of space—or polarization of the vacuum—and is responsible for so many of the special properties of the electric field. It does not have an analog in other fields like fluid dynamics or heat flow or diffusion because displacement current is ethereal and flows in empty space, in a vacuum. Mass cannot flow, conduct heat, or diffuse in a vacuum because mass is not present in a vacuum.

The magnetic field is created by flux $\mathbf{J}_{dielectric} + \mathbf{J}_{permanent} + \mathbf{J}_{other}$ and the displacement current $\varepsilon_0\, \partial \mathbf{E}/\partial t$. The fluxes describe the movement of the charges $\mathbf{Q}_{perm}$, $\mathbf{Q}_{dielectric}$, and $\mathbf{Q}_{other}$ of Maxwell's first law. Note that flux is not current nor is it proportional to the total current $\mathbf{J}_{total}$ of eq. (24) used in this paper.

The relation of charges and fluxes is determined by a combination of Maxwell's equations and the conservation of mass, continuity equations eq. (21)-(23), and constitutive equations that describe the properties of matter and charge. An example would be the Navier-Stokes equations [76, 84, 85, 97, 166-179] extended to deal with the flow of permanent charges. Another would be the Poisson-Nernst-Planck PNP [93-100] or drift diffusion equations [39, 43, 51, 101-106].

**Continuity equations describe accumulation** of that which flows (mass, charge, or current). When the time derivative (on the right-hand side) of the continuity equation is zero, the quantity that flows is conserved. When the time derivative of the continuity equation is not zero, the quantity that flows is not conserved. It accumulates. Only the total current $\mathbf{J}_{total}$ defined later in eq. (24) is always conserved within a system. The total current $\mathbf{J}_{total}$ never accumulates (ever, anywhere, under any conditions described by the Maxwell equations). The time derivative $\partial\, \mathbf{J}_{total}/\partial t = 0$ within an isolated system. In an un-isolated, open system, boundary conditions can make $\mathbf{J}_{total}$ vary with time, as can energy sources not included in the Maxwell equations at all, like convection or diffusion.

$\partial\, \mathbf{J}_{total}/\partial t$ equals zero in an isolated system because the components of $\mathbf{J}_{total}$ rearrange themselves (according to the Maxwell equations) so that $\mathbf{J}_{total}$ does not vary with time. An example is the EMF given in eq. (29). This rearrangement is particular striking in a series circuit where currents are equal at all times and in all conditions in the series elements of the circuit no matter how different is the microphysics of current conduction in each circuit element.

### Current in an Ideal Dielectric

$$\mathbf{J}_{dielectric} = (\varepsilon_r - 1)\varepsilon_0\, \partial \mathbf{E}/\partial t \tag{15}$$

$\mathbf{J}_{dielectric}$ is the current in classical perfect dielectrics with dielectric constant $\varepsilon_r$. $\mathbf{J}_{permanent}$ is the flux of mass with a fixed permanent charge. $\mathbf{J}_{other}$ includes charge movement produced by deformation of the shape of the mass; time dependence of polarization and induced charge beyond that of a perfect dielectric; charge that depends nonlinearly on the field; charge that depends on global, not local fields; and charge moved by other fields. Of course, all three currents— $J_{dielectric}$, $J_{permanent}$, and $J_{other}$—must be known either by experiments or constitutive field equations before Maxwell's equations can be usefully applied to compute forces, flows and energy in practical problems. Only currents specified by eq. (25) can be determined without such knowledge.

We separate current of a hypothetical ideal dielectric $J_{dielectric}$ from displacement current produced by polarization of space $\varepsilon_0 \, \partial \mathbf{E}/\partial t$. Our practice here differs from that of some textbooks because the two kinds of polarization are so different. One is a particular, highly complex and variable property of matter; the other is a universal property of space as stated so well, so clearly, so long ago by one Maxwellian, Jeans [141], p. 155 and p. 525.

We turn now towards the 'new' result in this paper, the conservation of total current $J_{total}$, which is shown to be true under all conditions that the Maxwell equations are valid. This is a more general result than found in textbooks, including Feynman [120], as far as I know. The result is a direct consequence of the properties of the vector operators divergence and curl: the divergence of the curl is zero whenever Maxwell's equations can be used, as can be verified by performing the vector operations or consulting the general theory of vector calculus [159-161].

We then have **conservation of current** discussed at length below, in more explicit modern form. If we separate the current $J_{dielectric}$ of an ideal dielectric, we can write

$$\mathbf{div}\left(\frac{1}{\mu_0}\mathbf{curl\ B}\right) \;=\; 0 \;=\; \mathbf{div}\left(J_{dielectric} + J_{permanent} + J_{other} + \varepsilon_0 \frac{\partial \mathbf{E}}{\partial t}\right) \qquad (16)$$

Traditional forms of **conservation of current** write the ideal dielectric as $\varepsilon_r \varepsilon_0 \, \partial E/\partial t$.

$$\mathbf{div}\left(\frac{1}{\mu_0}\mathbf{curl\ B}\right) \;=\; 0 \;=\; \mathbf{div}\left(J_{permanent} + J_{other} + \varepsilon_r \varepsilon_0 \frac{\partial \mathbf{E}}{\partial t}\right) \qquad (17)$$

**Derivation of Continuity Equation**. We derive the continuity equation relating flux/current and the storage of charge, providing a definition of capacitance as general as Maxwell's equations themselves, if one wishes, from eq. (21). Maxwell used the general idea of capacitance extensively in his reasoning, under the name 'Specific Inductive Capacity' [2] and Wolfgang Nonner showed me how the idea could be used to qualitatively understand the coupling (in a protein) of remote permanent charges (e.g., on aspartates buried in nonpolar regions of the protein) to ions moving in a channel in the protein.

Write **Ampere's law** in the representation used in eq. (17),

$$\frac{1}{\mu_0} \text{curl } \mathbf{B} = \mathbf{J}_{permanent} + \mathbf{J}_{other} + \varepsilon_r \varepsilon_0 \frac{\partial \mathbf{E}}{\partial t} \tag{18}$$

Take the divergence of both sides, recognizing that the divergence of a curl is zero whenever those operators are defined [159-161].

$$\text{div}\left(\frac{1}{\mu_0} \text{curl } \mathbf{B}\right) = \text{div}\left(\mathbf{J}_{permanent} + \mathbf{J}_{other} + \varepsilon_r \varepsilon_0 \frac{\partial \mathbf{E}}{\partial t}\right) = 0 \tag{19}$$

Solve the right hand equation in eq. (19) for the divergence of the current, interchange spatial and temporal differentiation, and use Maxwell's first equation (3) to derive

$$\text{div}(\mathbf{J}_{permanent} + \mathbf{J}_{other}) = -\text{div}(\varepsilon_0 \varepsilon_r \mathbf{E}) = -\frac{\partial}{\partial t} \text{div } \mathbf{D} = -\frac{\partial \boldsymbol{\rho}_f}{\partial t} \tag{20}$$

and get the (nearly) classical form.

### Continuity Equation

$$\boxed{\text{div}(\mathbf{J}_{permanent} + \mathbf{J}_{other}) = -\frac{\partial \boldsymbol{\rho}_f}{\partial t}} \tag{21}$$

$\mathbf{J}_{other}$ includes the nonideal components of current in real dielectrics; all other types of polarization; currents driven by other fields; nonlinearities and globally dependent current; and everything else. $\mathbf{J}_{permanent}$ is the current carried by permanent charge. Of course, all currents and charges must be known either by experiments or constitutive field equations before the continuity equation or Maxwell's equations can be usefully applied to compute forces, flows and energy in practical problems. Only currents specified by eq. (25) can be determined without such knowledge.

**Other representations of the continuity equation** that explicitly display the different types of charge and current may be useful.

Start with the right hand equation from the more explicit form shown in eq. (16) and substitute from eq. (12) to display the different types of charge $\mathbf{Q}_{perm}$, $\mathbf{Q}_{dielectric}$ and $\mathbf{Q}_{other}$. This yields another representation of the **continuity equation**.

$$\mathbf{div}\,(\mathbf{J}_{dielectric} + \mathbf{J}_{permanent} + \mathbf{J}_{other})$$
$$= -\frac{\partial}{\partial t}\Big(\mathbf{Q}_{dielectric}(x,y,z|t;\mathbf{E}) + \mathbf{Q}_{perm}(x,y,z|t) + \mathbf{Q}_{other}(x,y,z|t;\mathbf{E})\Big) \quad (22)$$

We reiterate (to avoid possible confusion) that $\mathbf{J}_{dielectric}$ and $\mathbf{Q}_{dielectric}$ describe the properties of ideal dielectrics, not the dielectric properties of real materials.

We use eq. (11) to write out $\mathbf{Q}_{dielectric}$ and interchange spatial and temporal differentiation to get still another form of the **continuity equation**

$$\mathbf{div}(\mathbf{J}_{permanent} + \mathbf{J}_{other} + \mathbf{J}_{dielectric})$$
$$= -\frac{\partial}{\partial t}\Big(\mathbf{Q}_{perm}(x,y,z|t) + \mathbf{Q}_{other}(x,y,z|t;\mathbf{E})\Big) - (\varepsilon_r - 1)\varepsilon_0\,\mathbf{div}\,\frac{\partial \mathbf{E}}{\partial t} \quad (23)$$

Note that non-ideal properties are described by $\mathbf{J}_{other}$ and $\mathbf{Q}_{other}$. The ideal properties are isolated and displayed in $\mathbf{J}_{dielectric}$ and the classical term $-(\varepsilon_r - 1)\varepsilon_0\,\mathbf{div}\,(\partial\mathbf{E}/\partial t)$.

The (nonzero) time rate of change of permanent charge $\partial \mathbf{Q}_{perm}/\partial t \neq 0$ in eq. (23) might seem paradoxical given the adjective 'permanent'. How does something permanent change? The answer is that it moves. The location of permanent charge is not fixed, even though its value is fixed. Permanent charge can move and flow, and thus its density at one location can vary in time, in the Eulerian coordinates we use here [70, 211, 212]. That is why I prefer the name 'permanent' charge to the more traditional 'fixed' charge.

Next, we define total current and show how it is conserved. It is useful to define total current $\mathbf{J}_{total}$ as Maxwell did, as described by his successors at Trinity Cambridge, Jeans and Whittaker: ref. [141], Ch. 17, p.511, and [7], p. 280, respectively.

## Definition of Total Current

$$\mathbf{J}_{total} = \mathbf{J}_{dielectric} + \mathbf{J}_{permanent} + \mathbf{J}_{other} + \varepsilon_0 \frac{\partial \mathbf{E}}{\partial t} \quad (24)$$

## Conservation of Current

$$\boxed{\mathbf{div}\, \mathbf{J}_{total} = 0} \quad (25)$$

$$\mathbf{div}\left(\mathbf{J}_{dielectric} + \mathbf{J}_{permanent} + \mathbf{J}_{other} + \varepsilon_0 \frac{\partial \mathbf{E}}{\partial t}\right) = 0 \quad (26)$$

**Continuity equations in general** describe how flows accumulate. Continuity equations show how flows are not perfectly conserved. If multiple fields are involved in flows, like convection and diffusion, for example, each field will have its own continuity equation and all continuity equations must be solved together, just as all the fields must be solved together, because they all interact. Variational methods are helpful in ensuring that all fields are solved together, consistently, with one set of parameters.

Note that the continuity equation for mass is not the same as the continuity equation for charge. It may seem surprising that the continuity equation for (the flux of) charge does not imply (by itself) the continuity equation for current. The current $\mathbf{J}_{total}$ defined here in eq. (24) includes the universal extra term, the ethereal current $\varepsilon_0\, \partial \mathbf{E}/\partial t$ that is an expression of the relativistic invariance of charge with velocity [48, 65, 126, 127, 180-183] and so flows in a vacuum and everywhere else. The ethereal current is a property of space, not a property of mass or charge.

Continuity equations for most flows—for flows of mass or charge, for example— allow accumulation because they include a time derivative (on the right-hand side). However, the time derivative does not appear on the right hand side of eq. (25). Current $\mathbf{J}_{total}$ as defined here and by Maxwell never accumulates. It is conserved perfectly, as long as the Maxwell equations apply. To say it again, $\mathbf{div}\, \mathbf{J}_{total} = 0$.

**Derivation of Conservation of Current.** So far, most of what we have written requires knowledge of charges and the idealized dielectric to be useful, The need to know the properties of $\mathbf{Q}_{perm}$, $\mathbf{Q}_{other}$ and $\mathbf{Q}_{dielectric}$ is an example of Feynman's instruction "… to exhibit in every case all the charges, whatever their origin, [so] the equations are always correct." [120]. But knowledge of those charges is not needed to derive conservation of total current $\mathbf{J}_{total}$ eq. (25). That derivation only depends on the properties of the divergence and curl operators and does not involve the properties of matter at all. (See derivation of eq. (16).)

The coupling of the electric and magnetic fields, including the universal polarization of the vacuum, that creates the ethereal current $\varepsilon_0\, \partial \mathbf{E}/\partial t$, allows us to violate Feynman's instruction: eq. (26) is remarkably powerful even in the face of complexity and ignorance of the properties of matter. We do not need to know the charges to derive conservation of the total current $\mathbf{J}_{total}$. Total current is perfectly conserved **_independent of any property of matter._** The derivation of equation (26) does not involve polarization of matter at all. It depends only on the polarization of the vacuum $\varepsilon_0\, \partial \mathbf{E}/\partial t$. Conservation of current is true inside atoms [60], in fact wherever the Bohm version of quantum mechanics can be applied [142-146].

This result is not just an abstraction. It is very important in practical applications as discussed on p. 8; in publications [60, 122-125]; and at the end of this paper. Indeed, I believe the universal nature of conservation of current is what allows the integrated circuits (of the computer being used to display or print this paper) to function robustly on the time scale of $10^{-10}$ sec and length scale of $10^{-8}$ meters.

**Solving for $\mathbf{E}(x, y, z|t)$.** We now determine the electric field that produces conservation of total current. We can directly integrate Maxwell's Ampere's law (14) or we can solve eq. (26) in general using the Helmholtz decomposition theorem [161, 213], as Chun Liu, taught me, with the same result.

$$\mathbf{E}(x,y,z|t) = \frac{1}{\varepsilon_0} \int_0^t \left( \frac{1}{\mu_0} \mathbf{curl\ B} - \left(\mathbf{J}_{dielectric} + \mathbf{J}_{permanent} + \mathbf{J}_{other}\right)\right) dt' \qquad (27)$$

In general, the **curl B** term in eq. (27) must be determined from the Maxwell equations and boundary conditions of a particular setup, a formidable task. But this is not necessary in the one-dimensional branched systems that define electrical circuits. **curl B = 0** in electrical circuits, as can be verified by direct substation in its definition.

For **electrical circuits,** because they are branched one dimensional networks

$$\text{div } \mathbf{J}_{total} = 0 \text{ implies } \mathbf{J}_{total} = 0, \quad \text{in circuits with zero initial conditions} \quad (28)$$

with $\quad \mathbf{E}(x,y,z|t) = \frac{1}{\varepsilon_0} \int_0^t \left(-\left(\mathbf{J}_{dielectric} + \mathbf{J}_{permanent} + \mathbf{J}_{other}\right)\right) dt'$

Thus, in electrical circuits, Maxwell's equations automatically provide the electric field of eq. (27) necessary to conserve total current, for any mechanism of the conduction of charge, independent of the magnetic field, because **curl B** $= 0$ in one dimensional systems like circuits. In my opinion, this special property is what makes our integrated circuits robust and reasonably easy to design using classical approaches originally derived for (nearly) **DC** macroscopic systems in the 1800's, even though our circuits work on time scales $10^8$ times faster and length scales $10^5$ times smaller than could be used in 1900.

Taking this approach, Kirchhoff's current law can be exact and universal, as shown with examples in [124, 125, 147]. Kirchhoff's law need not be the low frequency approximation described in textbooks and the literature [152-155]. The electric field, and polarization of space $\varepsilon_0 \, \partial \mathbf{E}/\partial t$, can have any value, and the polarization and conduction properties of matter are irrelevant as long as the system is nearly one dimensional. Kirchhoff's law applies only to one dimensional systems because it requires small **curl B** (see eq. (27)).

**One dimensional systems are not trivial.** One dimensional systems may seem trivial to physicists and mathematicians working in three dimensional systems, but they are no more trivial than our digital computers or smartphones. The circuits of our digital computers are nearly all branched one dimensional circuits designed with Kirchhoff's current law, more than anything else [33, 44, 45, 47, 148-151]. Digital computers can store and manipulate almost anything mankind has done, from dreams and images, to ideas, theorems, and computer programs, operating in $< 10^{-9}$ sec [33-50]. Remember that light travels approximately one foot in one nanosecond, so one dimensional branched systems that operate precisely and reliably at these speeds are not trivial. The validity of Kirchhoff's law, like the performance of electronic circuits at these speeds, is a welcome surprise, given textbook treatments—and the usual derivations [152-155]—of Kirchoff's law as a long time approximation.

I suspect that electrical circuits (i.e., branched one dimensional arrangements of components connected by wires) are used to implement high speed electronic technology because they allow design of robust devices with classical methods using Kirchhoff's current law [125, 147] without reference to the magnetic field.

**Speculation about circuit layout.** Circuits are idealizations of the three-dimensional arrangement of components in real integrated circuits. The layout of the circuit (i.e., its implementation in two and three dimensions) has important effects on the performance of real integrated circuits. The layout of the ground planes is particularly important, although ignored in elementary circuit treatments and discussed gingerly if at all even in advanced texts. Careful design of ground planes are needed to implement the simple point grounds of one-dimensional branched circuits [44, 45, 148-151] if integrated circuits are to function robustly at high speed with minimal cross talk and dissipation.

It would be interesting to examine actual layouts of integrated circuits to see how they control **curl B** [35, 46, 47, 148-151]. I suspect that high speed circuit boards are designed to minimize **curl B**. When $1/\mu_0$ **curl B** is significant (compared to $\mathbf{J}_{dielectric} + \mathbf{J}_{permanent} + \mathbf{J}_{other}$, of eq. (27)), current flow in circuits will not follow Kirchhoff's current law, because then $1/\mu_0$ **curl B** creates magnetic 'leakage', cross talk and dissipation as the magnetic field interacts with other materials that happen to lie nearby and are not in the (idealized) circuit diagram of the designed device. The (curl of the) magnetic field creates crosstalk that complicates design, to put it mildly. Reducing magnetic leakage and cross talk allows the real three-dimensional circuit to be well approximated by its one dimensional reduction, the idealized circuits of Kirchhoff's current law [125, 147].

**Electromotive Force EMF.** The electric field of eq. (27) can be used in electric circuits as a precise definition of Faraday's 'electromotive force' EMF [7] that moves charge and creates current.

$$\mathbf{E}(x,y,z|t) = \mathsf{EMF}$$
$$= -\frac{1}{\varepsilon_0} \int_0^t \Big( \mathbf{J}_{dielectric}(\cdots|t';\mathbf{E}) + \mathbf{J}_{permanent}(\cdots|t';\mathbf{E}) + \mathbf{J}_{other}(\cdots|t';\mathbf{E}) \Big) dt' \qquad (29)$$

Eq. (29) gives a modern and precise definition of an idea—EMF of circuits—that previously might have seemed mysterious or vague to some of us.

**Conservation of Current in a Series Circuit.** The importance (and power) of the conservation equation $\mathbf{div\,J}_{total} = 0$ is clearest [124] in simple series circuits [46-48, 163]. The conservation equation for $\mathbf{J}_{total}$ guarantees that all currents are equal in a series system even though the physics of conduction can be entirely different in the different components of the series system.

Series circuits are found 'everywhere' as parts of our electronic technology [33-50]. In those simple series circuits, the conservation equation $\mathbf{div\,J}_{total} = 0$ implies exact equality[5] of currents in every element at every time, **no matter what is the physics of the conduction of charge**.

In series circuits, current $\mathbf{J}_{total}$ is a practical reality that can be measured simply by inserting a low value resistor (chosen so it does not perturb anything of interest) in series with the other components. Exactly the same current will be measured no matter where the resistor is placed in the series circuit. Inserting the low value resistance provides an easy experimental test of the universal and exact nature of conservation of current in circuits.

The physics here is surprising, at least it was to me, when I thought it through in a practical example, shown in Fig. 3 of [124]. Consider a capacitor in series with a wire and a transistor. The microphysics of the capacitor is that of (say for discussion) a perfect dielectric involving the small reversible movements of charges; the physics of the wire is delocalized electrons producing a substantial electromagnetic field outside the wire at times shorter than say $10^{-6}$ sec; the physics in the transistor is that of the drift diffusion of holes and electrons [39, 40, 43, 104, 214-216] (which are quasi-particles, not the particles observed in cathode rays by Thomson [8]). **How can currents be equal when they arise in such different ways?** If the capacitor in series with the semiconductor is changed, the current in the semiconductor is changed, and *vice versa*.

How does this happen? How does the microphysics of the semiconductor know about the change in the capacitor? How do the forces on holes and electrons change when the capacitor is changed?

The answer is that the electric field $\mathbf{E}(x, y, z|t) =$ EMF of the circuit—determined by solving eq.(24)—changes the currents of eq. (26) by exactly the amounts necessary to make the total current $\mathbf{J}_{total}$ exactly equal in every element of a series circuit at every time and in every condition.

---

[5] 'Exactly the same' means that current is equal at any time, at any voltage or current, in any conditions, including when fluxes are driven by fields not included in the Maxwell equations at all, to the precision that current can be measured at all.

The electric field $\mathbf{E}(x,y,z|t)$ of eq. (29) changes the displacement current $\varepsilon_0\,\partial\mathbf{E}/\partial t$ and the individual fluxes $\mathbf{J}_{...}(\cdots;\mathbf{E})$ by just the amounts necessary to make conservation of total current $\mathbf{J}_{total}$ exact, everywhere at every time, under all conditions in series circuits.

The microphysics of conduction obviously cannot make all the currents the same, in itself.

The name 'microphysics' implies a physics on only the microscale, without macroscopic interactions. Physics on the microscale cannot enforce equality of current in a series system because the microscale ignores the other devices (in series) that also control current flow. A global term is needed to enforce equality of current in a series of devices and that term arises from polarization. It is present even in the absence of matter because of the ethereal term. ***Understanding a series circuit requires understanding of Maxwell's version of Ampere's law and its consequences,*** eq. (14), (21), (25) and (29).

The field specified by the Maxwell equations (for a series circuit) changes the individual currents $\mathbf{J}_{...}$ according to their constitutive laws and conservation of mass, as described in some cases by the Navier-Stokes equations [76, 84, 85, 97, 166-179] and others by the Poisson-Nernst-Planck, PNP [93-100] or drift diffusion equations [39, 43, 51, 101-106]. The change is different in every component and can be different at different places in the same component. But current is everywhere the same in series circuits even when fields quite different from electrodynamics are a source of current. Maxwell's equations are universal and exact even in the presence of convection and diffusion, and other fields.

**Everyday conservation of current: complete the circuit.** The physics of current conservation in the series circuit is less surprising perhaps in the context of our everyday life than in the abstract concept of series circuits. In our everyday life, everyone knows that current must be given a complete path. If the series path is broken, and the series circuit is interrupted, nothing happens. In mathematical language, if current is forced to be zero in one place (in a series circuit), it must be zero everywhere by eq. (28). In the everyday language of series circuits, an 'open circuit' stops current flow. A circuit must be 'closed' to work.[6]

**Importance of Computing the Field.** The conservation law eq. (25) does not involve the properties of matter. It is difficult to exaggerate the importance of this fact.

This mathematical fact implies unexpected physics which seems to differ from the expectation of Feynman [120] that knowledge of all charges is needed to implement Maxwell's equations in real systems. The implementation of Maxwell's equations in the circuits of our computers using Kirchhoff's current does not depend on the knowledge of charges or their polarization. It depends only on the polarization of the vacuum $\varepsilon_0\, \partial \mathbf{E}/\partial t$.

The derivation of conservation of current shows that Maxwell's equations *in themselves* will provide exactly the electric and magnetic fields necessary to ensure exact conservation of current for any physics of conduction of charge, for any geometry of conductors, for any properties of matter, at any time, and under any conditions in which Maxwell's equations are valid.

The field that produces conservation of current does two things in circuits. It creates the ethereal displacement current $\varepsilon_0 \partial \mathbf{E}/\partial t$ (eq. (4) necessary to produce conservation of current. The electromagnetic field also creates the forces that move charges with mass so they provide exactly the current $\mathbf{J}_{dielectric}(x,y,z|t;\mathbf{E}) + \mathbf{J}_{permanent}(x,y,z|t;\mathbf{E}) + \mathbf{J}_{other}(x,y,z|t;\mathbf{E})$ necessary to produce conservation of current. Both are described by the continuity eq. (21)-(23).

The currents $\mathbf{J}_{\ldots}(x,y,z|t;\mathbf{E})$ are produced by the flux of mass with charge. The charges involved in this flux are $\mathbf{Q}_{\ldots}(x,y,z|t;\mathbf{E})$ of eq. (22) and eq. (12), derived from Maxwell's second and first laws respectively. The charges $\mathbf{Q}_{\ldots}(x,y,z|t;\mathbf{E})$ move because of the forces on those charges. Those forces are determined by 'Coulomb's law' (the integral version of Maxwell's first law) and perhaps other fields like convection and diffusion, if they are present, along with the continuity eq. (21)-(23) and the continuity equations of mass that are part of field theories of convection and diffusion (for example).

---

[6] This circuit language was commonplace in the 1800's but I find it confuses some born in the 2000's: they want to know what is 'open' and what is 'closed'?

**Macroscopic Laws Move Atoms.** The forces defined in eq. (27) exist on all time and distance scales.

Maxwell's equations provide the electromagnetic fields that move atoms as needed to satisfy conservation of current. The electromotive force EMF of eq. (29) moves atoms individually as well as collectively. The **_apparently macroscopic conservation laws have effects on all scales_**. Maxwell's (apparently) macroscopic equations move individual atoms because the electric field moves atoms as well as objects.

Once Maxwell's equations are updated, there is no need to separate atomic scale and macroscopic scales. One theory will do because the equations are obeyed at all times and under all conditions.

Indeed, in the simplified geometries of circuits—that are one dimensional branched networks—we do not even need a theory of magnetism. Magnetic fields cannot divert energy and current, or create cross talk if they are not present. In circuits, the electric field $\mathbf{E}(x, y, z|t)$ **_itself_** links all scales exactly the right way to guarantee conservation of current in Kirchhoff's law. Perhaps that is why circuits controlling electric current can be designed with theories originally derived for low frequencies—nearly DC—even though they operate at times less than one nanosecond [33, 44, 45, 47, 148-151].

**Importance of displacement current.** The role of the ethereal displacement current $\varepsilon_0 \partial \mathbf{E}/\partial t$ of eq. (4) cannot be overstated. In a universe without this current, the electromagnetic field in a circuit would not be determined by an integral like eq. (29). Conservation of current would not be universal and exact, independent of matter, true on all scales. Conservation of current would not be enough to exactly implement the circuits of our computer technology.

It seems no coincidence that the ethereal current $\varepsilon_0 \partial \mathbf{E}/\partial t$ that creates the forces and fluxes that make conservation of current exact on all scales in circuits is also the term that allows electromagnetic waves to move through a vacuum. Indeed, this is the term that makes the charge on an electron independent of velocity even at velocities approaching the speed of light [48, 65, 126, 127, 180-183]. It is the ethereal polarization of the vacuum $\varepsilon_0 \partial \mathbf{E}/\partial t$ discovered by Maxwell that creates all these special properties of electrodynamics [7, 141].

**Ethereal polarization** $\varepsilon_0 \partial \mathbf{E}/\partial t$ links the properties of space and time as described by the theory of special relativity for inertial systems, that do not rotate [48, 180-183]. Rotating systems require some form of general relativity [65, 126, 127, 217].

**General implications** of our analysis have already been discussed in the section "Rewriting Maxwell's Equations", p. 8-11.

We now document how **classical Maxwell equations can be misleading** if used in their original form without permanent charge.

**Dissolved substances are permanent charges, not potentials.** First we consider ionic solutions, like the ~140 mM $Na^+Cl^-$ (extracellular) or $K^+Cl^-$ (intracellular) in which life occurs [133, 218].

It is difficult to describe ionic solutions, at all, if field equations are used that do not include permanent charge, like the classical Maxwell equations. More or less anything that dissolves in water has permanent charge [22, 27], from hard sphere ions to proteins. Indeed, most matter and chemical compounds and bonds [18] have permanent charge, often of high density [219].

In my view, treating proteins, solutes, or boundary conditions as distributions of potential, not permanent charge, has led to difficulties of some importance. The electric field valid under one set of conditions has been held constant and transferred mistakenly to other conditions in which Maxwell's equations (etc.) require the field to be different.

Under one set of conditions the solution of electrostatic problems is unique and so electric forces can be described in several ways.

(1) The electric field can be said to come from the surface charge of a protein. In that case, the electric field is described by the inhomogeneous Neumann problem derived in the Appendix of [137].

(2) The same electric field can be said to arise from surface potential on the protein. In that case, the electric field is described by an inhomogeneous Dirichlet problem. Both treatments give the same result under one set of conditions. The classical Maxwell equations give the correct relation between charge and field.

When systems are transferred from one set of conditions to another, results depend on the nature of the surface charge and so the classical Maxwell equations give incorrect results.

The representations (Dirichlet and Neumann) change in drastically different ways that are nothing like equivalent when experimental conditions are changed because the nature of charge is different in Dirichlet and Neumann representations.[7] The Neumann (permanent charge) representation is a natural representation for matter by itself, isolated from sources of energy or charge. The Dirichlet condition is an unnatural augmented representation because it requires connection to the outside. It describes a system that is ***not isolated*** but rather requires charge, energy, and mass from external sources (to maintain the fixed Dirichlet boundary condition as other conditions change).

The different properties of representations create serious problems unless handled explicitly, particularly where ionic solutions are involved. Experimental and technological applications almost always change conditions a great deal, either by changing the potential at a specific location using an external experimental apparatus that generates charge and uses energy ('voltage clamp', e.g., [220, 221]), or by changing the ionic composition or content of ionic solutions (e.g., [222]) and thereby changing the screening of charges.

**Screening in ionic solutions.** In fact, the dominant reality of ionic solutions is that properties are determined in large measure by screening (or shielding as it is sometimes called) of charges, as documented in textbooks of physical and electrochemistry, e.g., [17, 18, 20, 22, 23, 25-27, 30, 31, 223, 224]. The reality of shielding is seen dramatically in simulations of ionic channels, even those with only permanent dipolar charges, like the carbonyls of gramicidin [225].

As a general rule, when dealing with ionic systems, concentrations are changed, screening is changed and so ***computing the field is a necessity***, as discussed some time ago in the biophysical literature [128, 129], and known to the computational electronics community long before that [43, 103-105, 215, 226, 227].

The existence of shielding is easily shown experimentally in biology and electrochemistry, simply by adding background salt (e.g., $Na^+Cl^-$) to the solution, or other ions that are not involved in the transport or chemical reaction of interest. Models of ionic solutions that maintain fixed potential profiles (i.e., models which fix the potential as a function of location and/or time, e.g., the constant field models of physiology [130, 204, 222, 228]—that followed (K.S. Cole, personal communication, ~1961) Mott's constant field model of solid state 'crystal' rectifiers [229]—cannot deal with the phenomena of shielding, because the essential feature of shielding is the change in potential profile (with concentration of mobile ions) not present in models that assume constant fields.

---

[7] The problem is subtle and recurring in my experience, particularly, for those coming from a theoretical background, and not used to the changing conditions typical of experimental and technological science. Mathematicians for example are not used to studying the variation of boundary conditions, or the sensitivity to boundary conditions, unless they are explicitly asked to do so. Scientists often forget to ask.

Indeed, assuming a profile of the electric field (i.e., of the electric potential) as salt is changed is equivalent to assuming that the electrical forces do not change as the composition or contents of ionic solutions change. The composition of an ionic solution is its concentration of permanent charges of different types, i.e., of different chemical species of ions, like $K^+$ and $Na^+$. It is obviously silly to assume that electric forces do not change as the concentration of permanent charge changes.

In fact, the only way to maintain a fixed profile of electric forces as salts are changed is to inject charge at many places along that profile. This is rarely done in in biological or technological systems or applications. Indeed, maintaining a fixed profile of potential, independent of other variables, is surprisingly difficult even in apparatus built for that purpose using a `SQUID` (semiconductor quantum interference device) [230] because a handful of charges produces a large change in the profile of potential.

Injecting charge in the theory (when it is not injected in the experiment) is then seen as injecting an artifact. The artifact is likely to be large (estimated in the Appendix of [124]) given the enormous strength of the electric field. A small charge creates large forces, described as a large potential, as vividly explained on the first page of Feynman's textbook [120]: one percent of the charge in a human produces (at one meter distance) the force necessary to 'lift the earth'.

**Exponentially large artifact.** If flux or current flows over a large barrier, as it often does [204, 231, 232], an exponentially large artifact is likely to occur. Artifactual charge would be injected (in these models) at the top of large potential barriers, where flux is exponentially dependent on charge and potential. These problems have been discussed in the context of ionic channels in a series of papers [128-130, 233-236] that develop practical models not subject to these problems.

The potential profile must be computed from the charges, because the potential varies so much as experimental conditions change. Experimental conditions are likely to change screening a great deal. They are much less likely to change the permanent charge. If this statement seems problematic, it might be useful to actually compute and examine the sensitivity function defined as the derivative of the output of the system with respect to the condition being varied.

**Digression Concerning Sensitivity.** The explicit analysis and computation of sensitivity (with respect to various parameters) is found to be important in the theory of inverse problems, where ill-posedness is a central issue [237-241]. One might argue that much of science and even more of biology is an inverse problem [242] that seeks to find how evolution has created adaptations to solve problems that limit reproduction. These comments apply to ionic solutions in general, but they are particularly important when dealing with the macromolecules of life, chiefly proteins and nucleic acids.

**Proteins and surface permanent charge.** Returning to biological applications, we find that the proper description of charged surfaces is important to understanding proteins. Proteins are the robots of life [29, 136], responsible for a large fraction of biological function. Nucleic acids DNA and RNA (in its several forms) carry the genetic information that allows life to be inherited.

Proteins 'bristle with charge', in a saying attributed to Cohn and Edsall [243, 244] by Tanford [29, 136]. (Cohn, Edsall and Tanford were referring to the permanent negative and positive charges of acid and base side chains, glutamates E, aspartates D, arginines R and lysines K, more than anything else.) Proteins have large densities of permanent negative and positive charges. Concentrations of ions around 20 molar are found in locations important for protein function, namely ion channels, active sites of enzymes (catalytic active sites [245]), and binding sites, including drug binding sites. Nucleic acids have similar concentrations of mobile ions near their highly charged external surfaces. For reference, the concentration of *solid* $Na^+Cl^-$ is around 37 molar and *solid* $Ca^{2+}Cl_2^-$ is around 194 molar.

As a rule, ion concentrations tend to be very large where ions have important roles in devices [242], in biology and technology [168, 245-253].

**Inconsistent Theories of Thermal Motion.** Unfortunately, the classical theory of thermal (i.e., Brownian motion) is subject to a similar criticism [60, 236]. In these theories, forces are computed in ionic solutions assuming the electrical potential is a known function that does not fluctuate as concentrations of ions fluctuate. The electrical potential is simply a short hand for electrical forces and it is obvious that those forces must vary a great deal (p. 1-1 of [120]) as ion concentrations fluctuate in thermal motion. Simulations of molecular dynamics confirm the obvious.

The assumption of time independent potentials found in many theories of Brownian motion is inconsistent with the fundamental properties of the electric field. If the concentration of charge varies with time, the electric forces must vary as well. Approximations may be possible in special cases, but those approximations must be derived and computed. They must be checked, and errors shown to be reasonable.

The inconsistent treatment of electric forces seems widespread in the mathematical literature of Brownian motion and the physical literature of statistical physics. Perhaps some of the phenomena labelled as anomalous diffusion or anomalous Brownian motion arise from this inconsistency [254].

The importance of fluctuations in electric field in these anomalous phenomena is easy to check experimentally. Simply vary background salt and see if the anomalous phenomena vary as they should. Changing salt composition or concentration should have a large effect on some anomalies because changing salt changes the shielding of charges and thus changes the fluctuations in forces on ions often described by potentials.

Another major difficulty concerns conservation of current.

**Kinetic Models and Conservation of Current.** Chemistry arose historically as the science that changed one substance into another. These changes were naturally described by arrow models $A \rightarrow B$. These models were made quantitative by using the law of mass action, ascribing an equilibrium constant to the reaction, and forward and backwards rate constants to the forward and backwards components of the reaction. The charge and flow of charge in the reaction were overlooked altogether, or not a subject of much attention.

The unfortunate consequence of this oversight was that the currents in a sequence of reactions $A \rightarrow B \rightarrow C$ were not compared. $A \rightarrow B$ and then $B \rightarrow C$ were not compared. Maxwell's equations require the currents $\mathbf{I}_{A \rightarrow B}$ and $\mathbf{I}_{B \rightarrow C}$ to be equal as we have seen, because the reactions are in series. But the law of mass action of chemical kinetics does not require the currents to be equal. That law was designed to conserve mass, not current. Auxiliary conditions can be used to make these currents equal, but those conditions must involve the full range of solutions of Maxwell's equations, and probably the flow equations of matter, because electrodynamics and conservation of matter are global. What happens in one place changes what happens in another. Consideration of a simple series circuit makes this obvious. Everyone knows from everyday household experience that a complete circuit is needed for current flow. Interrupting a circuit anywhere, interrupts current flow everywhere.

Perhaps a network that satisfies conservation of current needs to be solved along with the classical network that satisfies conservation of mass, so the kinetic models are consistent with both conservation laws, using one set of unchanging parameters, so models are transferrable and useful in technology. The combination of two network models (for conservation of mass and also for conservation of current) might be solved with a variational principle in the spirit of Chun Liu's **EnVarA** [72, 74, 76, 79, 82-85, 90, 170, 209, 210].

The consequences of the violation of conservation of current in models of chemical kinetics are profound but we will not present details here since they have been extensively discussed in other publications [60, 122-124]. Numerical estimates of the errors that can occur are in Appendix of [124].

# Consequences of Updating Maxwell's Equations

Maxwell's original equations correctly describe the relation of charge and potential. Maxwell's equations can be updated to describe charge, its movement, and current, as we have seen eq. (4), (12), (14), and (20). With this update, they include the permanent charge of electrons, ions, and molecules that were unknown before 1897, to pick the date when the electron was discovered in beta rays in a (near) vacuum, and Maxwellians (mostly in England) were convinced of the existence of permanent charge.

Rewriting the equations is trivial, and even petty in one sense, as eq.(6)–(28) demonstrate, but profound in another, as p. 8-12 try to demonstrate.

Rewriting is an update that demonstrates the universal nature of Maxwell's equations entirely independent of the properties of matter and material polarization. With this more visible legitimacy, it becomes clear that Maxwell's equations need to be included in most models of mass transport (because most mass transport also involves charge transport and polarization). In particular, conservation of total current needs to be satisfied in a wide variety of systems where it has not traditionally received much attention.

Rewriting the equations emphasizes what is needed from scientists as they deal with charge in matter. Complete descriptions of charge and polarization are needed except when knowledge of total current eq. (25) is enough.

Updating Maxwell's equations to deal with charge on matter benefits from the separation of the properties of matter, and its flow, from the properties of space and the flow of current (for example, in a vacuum). The polarization of the vacuum (i.e., the polarization of space) is characterized by a universal and exact expression for the ethereal current $\varepsilon_0\, \partial \mathbf{E}/\partial t$ that is an unavoidable consequence of Maxwell's version of Ampere's law. The separation of current into its ethereal component $\varepsilon_0\, \partial \mathbf{E}/\partial t$, and the flow of material charge, allows the derivation of a universal and exact version of conservation of total current in circuits, that uses the definition of total current $\mathbf{J}_{total}$ Maxwell used.

The universal nature of conservation of total current of implies constraints necessary to make theories of Brownian motion and chemical kinetics satisfy the equations of electrodynamics, something they do not do in their classical or usual form.

The updated derivation of conservation of total current has profound effects on our understanding of the electrical circuits of our digital technology. In systems of components in series, conservation of total current becomes equality of current, something hard to understand on the nanosecond time scale, using just the classical Maxwell equations. Conservation of total current shows that Kirchhoff's current law (in branched one dimensional systems) is not the steady state approximation it is stated to be in most places in the literature, and in the minds of most scientists.

Conservation of total current shows how Kirchhoff's current law can apply from the time scale of the nano- (nearly pico-) switches in our technology to the macroscopic time scale of biological function and everyday life. Conservation of total current shows how ions in the wildly fluctuating, tightly crowded confines of a protein can be observed as they perform their natural function, injecting a 'gating current' across a cell membrane [255].

Some of the gating current will flow through the conduction pore of a voltage sensitive channel. That current might trigger the opening of the channel. In the same spirit, the binding of a transmitter (like acetylcholine or glutamate) to a charged binding site on a chemically gated channel might inject a current into its conduction pore and trigger its opening.

# Conclusion

Maxwell's equations have always described correctly the relation of charge and electric force. That model of the electric field is universal—as universal as anything known to science—as shown by the intimate connection of Maxwell's equations with special and general relativity.

It is no surprise that Maxwell's equations can be easily updated to describe the charge itself. What is surprising, at least to me, is that Maxwell's equations can be updated to describe conservation of total current, ***independent of any property of matter,*** with no approximation beyond those inherent in the equations themselves.

It seems that all theories and simulations of matter that involve electricity must satisfy conservation of total current, even if that requires significant modification of traditional formulations. Otherwise, artifactual charges accumulate, errors in potential can be large (Appendix: [124]): the electric field is extraordinarily strong, beyond our intuition, as Feynman told us in the beginning, on the first page of [120], so artifactual charges produce large errors in potential. Fluxes over barriers of potential can depend exponentially on barrier height, so artifactual charges can produce exponentially large errors in flux.

# **Acknowledgement**

Chun Liu made me understand the need for a **curl** term in eq. (27). Weishi Liu and Carl Gardner had tried previously but I had not understood. The idea that current injection could open a channel arose in most productive (and joyful) discussions with Manuel Landstorfer. Uwe Hollerbach helped remove errors. Many thanks to them.

This paper arose from discussions with Ardyth Eisenberg, who joined me in choosing the title and made many other contributions beyond her incomparable editing. It is a professional responsibility to recognize all she has done for this paper. It is a personal pleasure to thank her for all she has done for the paper, for me, and for our family.